\documentclass[aps,prl,reprint,nobibnotes, floatfix,
longbibliography,superscriptaddress]{revtex4-2}
\usepackage{t1enc}
\usepackage[utf8]{inputenc}
\usepackage{amsmath}
\usepackage{graphicx}
\usepackage{siunitx}
\usepackage{textcomp}
\usepackage{comment}
\usepackage{float}

\begin{document}

\title{Millikelvin Si-MOSFETs for Quantum Electronics}

\author{Nikolai Yurttagül}\email{nikolai.yurttagul@semiqon.tech}\affiliation{SemiQon Technologies, Tietotie 3, 02150 Espoo, Finland}
\author{Markku Kainlauri}\affiliation{SemiQon Technologies, Tietotie 3, 02150 Espoo, Finland}
\author{Jan Toivonen}\affiliation{SemiQon Technologies, Tietotie 3, 02150 Espoo, Finland}
\author{Sushan Khadka}\affiliation{SemiQon Technologies, Tietotie 3, 02150 Espoo, Finland}
\author{Antti Kanniainen}\affiliation{University of Jyväskylä, NanoScience Center, Department of Physics, PB 35, 40014, Jyväskylä, Finland}
\author{Arvind S. Kumar}\affiliation{University of Jyväskylä, NanoScience Center, Department of Physics, PB 35, 40014, Jyväskylä, Finland}
\author{Diego Subero}\affiliation{SemiQon Technologies, Tietotie 3, 02150 Espoo, Finland}
\author{Juha T. Muhonen}\affiliation{University of Jyväskylä, NanoScience Center, Department of Physics, PB 35, 40014, Jyväskylä, Finland}
\author{Mika Prunnila}\affiliation{SemiQon Technologies, Tietotie 3, 02150 Espoo, Finland}
\author{Janne S. Lehtinen}\email{janne.lehtinen@semiqon.tech}\affiliation{SemiQon Technologies, Tietotie 3, 02150 Espoo, Finland}

\DeclareSIUnit{\nothing}{\relax}


\begin{abstract}
\noindent Large power consumption of silicon CMOS electronics is a challenge in very-large-scale integrated circuits and a major roadblock to fault-tolerant quantum computation. Matching the power dissipation of Si-MOSFETs to the thermal budget at deep cryogenic temperatures, below 1 K, requires switching performance beyond levels facilitated by currently available CMOS technologies. We have manufactured fully depleted silicon-on-insulator MOSFETs tailored for overcoming the power dissipation barrier towards sub-1 K applications. With these cryo-optimized transistors we achieve a major milestone of reaching subthreshold swing of 0.3 mV/dec at 420 mK, thereby enabling very-large-scale integration of cryo-CMOS electronics for ultra-low temperature applications.
\end{abstract}

\maketitle

\noindent Cryogenic CMOS holds great potential for emerging fields such as high-performance computing \cite{SALIGRAM2024100082,9431559,9265065}, space technologies \cite{Saint-Pe2005,9265065}, and quantum nanoelectronics \cite{Xue2021,9567747,Pauka2021,Gonzalez-Zalba2021,Oka2022,Underwood2024,Bartee2025}. Fault-tolerant quantum computation is projected to require low-noise connectivity and electronic control of up to millions of qubits at cryogenic temperatures \cite{De_Michielis_2023}, which is achievable by operating Silicon CMOS integrated circuits at ultralow temperatures. The integration can be heterogeneous (Fig. 1a), which offers flexibility between different technologies and qubit modalities, or monolithic (Fig. 1b), which reduces parasitic losses and enables higher packaging density. \\

\noindent The use of cryogenic CMOS has been limited by high power losses compared to available thermal budgets. Established and scalable qubit platforms typically operate below 1 K, where current cryo-CMOS power metrics would overwhelm the cooling power of available refrigeration methods, usually in the range of a few $\si{\micro\nothing}$W to several mW \cite{Pobell:2007egf}. Scaling down power dissipation of CMOS electronics accordingly requires reducing both static power loss, by reducing off-state leakage currents, as well as dynamic power loss, by reducing the supply voltage $V_\mathrm{dd}$ \cite{TEWKSBURY1985255}. The ability to reduce both simultaneously is limited by the diffusive subthreshold conductance of a MOSFET, quantified by the subthreshold swing $SS = \mathrm{d} V_g/\mathrm{d}(\mathrm{log_{10}}I_d)$, with the drain current $I_d$ and the gate-voltage $V_g$ \cite{Lundstrom}. The lower limit of $SS$ is set by the extent of thermal excitations of conduction electrons at temperature $T$, approximated by the lower thermionic limit $SS(T)=\mathrm{ln}(10) k_{\text{B}} T/e$, with the Boltzmann constant $k_{\text{B}}$, and elementary charge $e$.\\

\begin{figure}[b]
\includegraphics[width=0.45\textwidth]{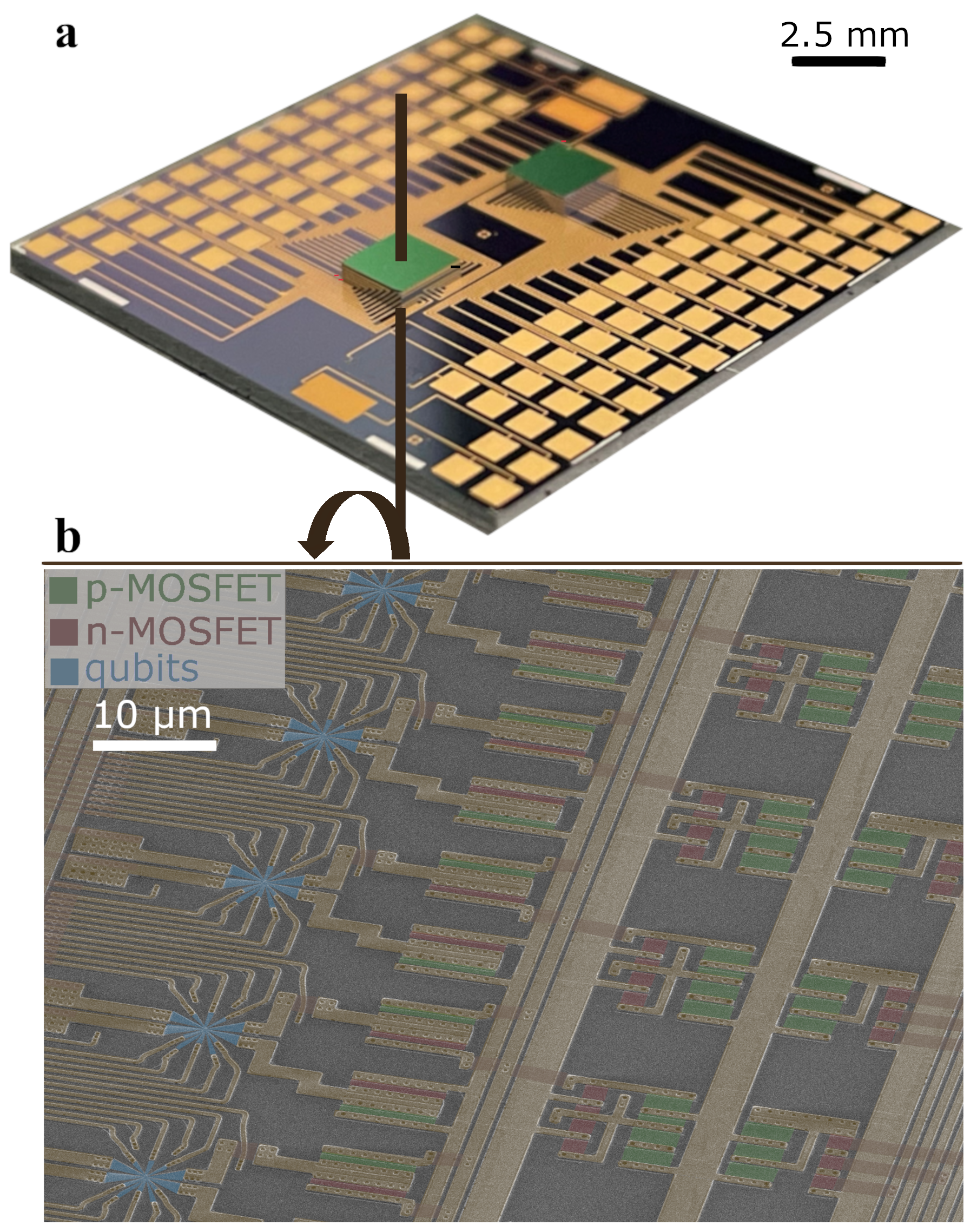}
\caption{Examples of heterogeneous and monolithic integration. \textbf{a} Photograph of a flip-chip assembly of hybrid classical and quantum chiplets, bonded to a I/O board. \textbf{b} False colored SEM micrograph of a planar 
 cryogenic quantum IC, that could be used as a chiplet in \textbf{a},  containing SiMOS qubit devices and digital CMOS logic blocks for signal demultiplexing and control. MOSFET devices from the IC blocks, highlighted in green and red, are characterized in this work with respect to switching metrics}
\end{figure}

\noindent However, the thermionic limit of switching is not reached in real MOSFETs, as subgap states, which start to dominate the subthreshold current below a characteristic limit temperature, cause a freeze-out of the sub-threshold conductance \cite{8946710,8660508,8326483,8329135,8946710,8066591,Balestra_2017}. This effect becomes more significant towards ultra-low temperatures, imposing an unfavorable scaling for operating cryo-CMOS electronics. Around liquid $^4$He temperatures (4.2 K), measured values for SS are usually well above 10 mV/dec for a wide variety of MOSFET types, which is an order of magnitude larger than the thermionic limit $SS(\text{4 K})=0.8$ mV/dec. The lowest reported values for foundry-fabricated FD-SOI MOSFETs are not lower than 5-7 mV/dec at 4 K \cite{9265034,8660508} and further cooling towards millikelvin temperatures has not shown significant improvement so far \cite{8660508,8326483,8329135,9567747,9567802,Balestra_2017,8066591,8946710}, except for a recent study \cite{Oka2023} where a further drop of SS was found below 100 mK after initial saturation.

\begin{figure*}[t!]
\centering
\includegraphics[clip,trim=0.6cm 7.6cm 0.2cm 5.5cm,width=0.95\textwidth]{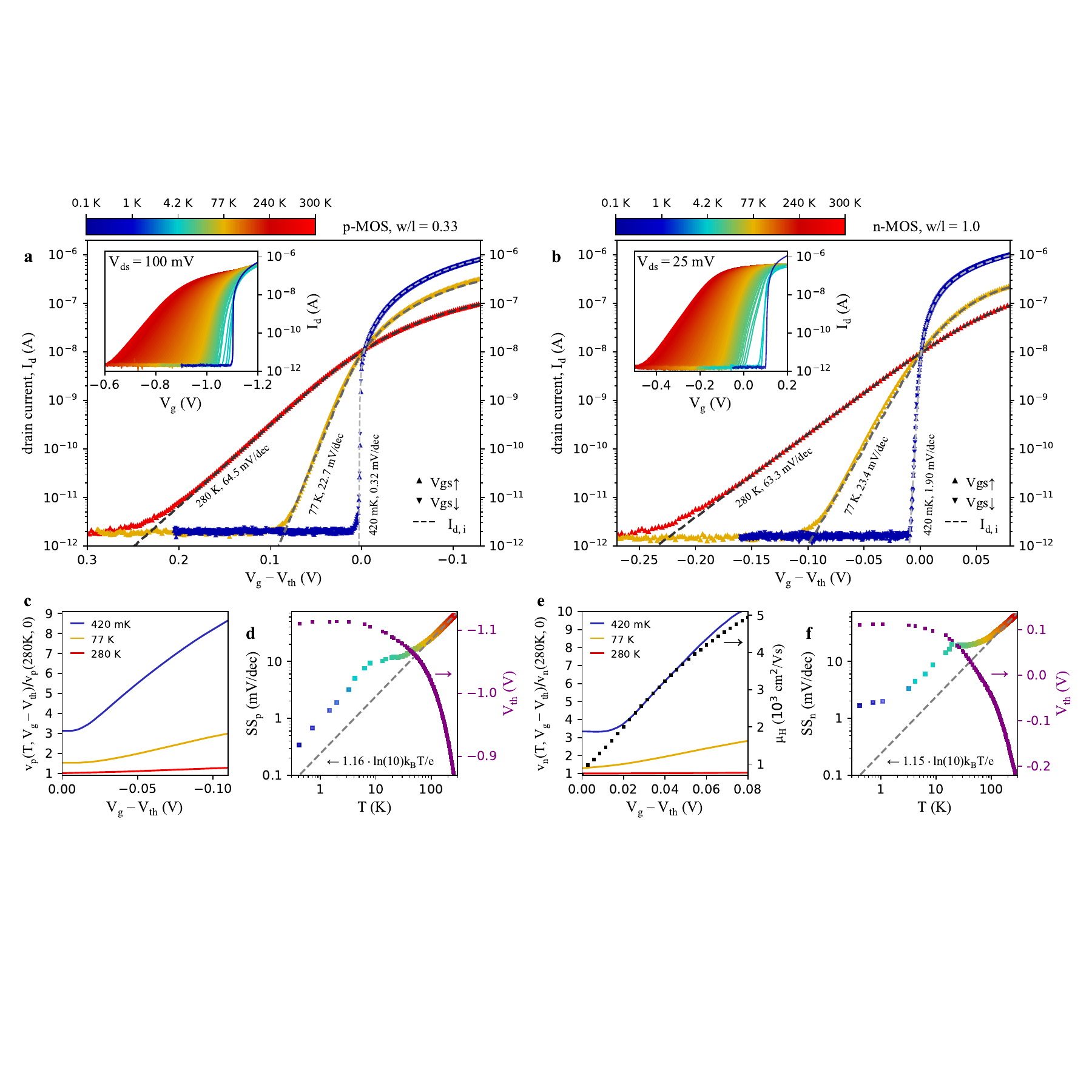}
\caption{FD-SOI MOSFET transport at variable temperature, $I_d(V_g)$ transfer characteristics for a p-MOSFET (\textbf{a}, $V_{ds}=100$~mV), and an n-MOSFET (\textbf{b}, $V_{ds}=25$ mV) between 280~K and 420~mK. The arrows in the legend indicate the sweep direction of $V_g$ (see also Fig. 3 for details). Raw data is plotted in the inset panels and transfer characteristics at 420~mK, 77~K, and 280~K is highlighted in the main panel. The curves in the main panel are fitted to Eq.~3 ($J_\mathrm{ds,i}$, arb. unit) to extract $SS$, and $v_i$. \textbf{c,e}: $v_i(V_g)$ above threshold, all values are normalized to the subthreshold value at 280~K. In \textbf{e} the measured mobility for a n-Hall-bar with the same $C_\text{ox}$ is plotted for comparison. \textbf{d,f}: $SS(T)$ and $V_\mathrm{th}(T)$ plotted against temperature. Adjusted thermionic limits $SS_i(T)=s_i\mathrm{ln}(10) k_{B} T/e$ are plotted with dashed lines for comparison.}
\end{figure*}

\noindent Apart from a sharper turn-on, other key MOSFET and circuit metrics improve towards lower temperatures, such as reduced wire delays, increased on-current due to increased carrier mobility, lower gate leakage currents, and higher charge retention times. Challenges originate from quantum effects \cite{9777765,9096333,BONEN2022108343} and a thermal shift in threshold voltage \cite{Oka2022}. The latter can be controlled in FD-SOI MOSFETs, making it an attractive technological platform for cryo-CMOS \cite{9265034}. Additionally, in fully depleted channels one can profit from a high mobility and possibly a reduced localization of charge carriers \cite{Shashkin-2019}.

\noindent Accounting for this, we experimentally evaluate FD-SOI MOSFETs, fabricated on a custom CMOS pilot-line. The gate stack fabrication is tailored to sub-1 K applications, such that a low impurity density facilitates low charge noise. For these gate-stacks a combination of very low 1/f noise with record low $SS$ of 4 mV/dec at 5.6 K was already demonstrated earlier, consistent over a range of physical gate length from 50 nm to several \si{\micro\nothing}m  \cite{bohuslavskyi2024scalableonchipmultiplexingsilicon}. In this work we evaluate the switching behaviour of our cryo-MOSFETs down to 420 mK and show by measuring gate transfer characteristics that $\sim$1 mV/dec switching in Si MOSFETs is possible at millikelvin temperatures, enabling ultra-low power operation of millikelvin cryo-CMOS.\\


\begin{figure*}[t!]
\includegraphics[clip,trim=0.62cm 7.8cm 0cm 5.7cm,width=0.95\textwidth]{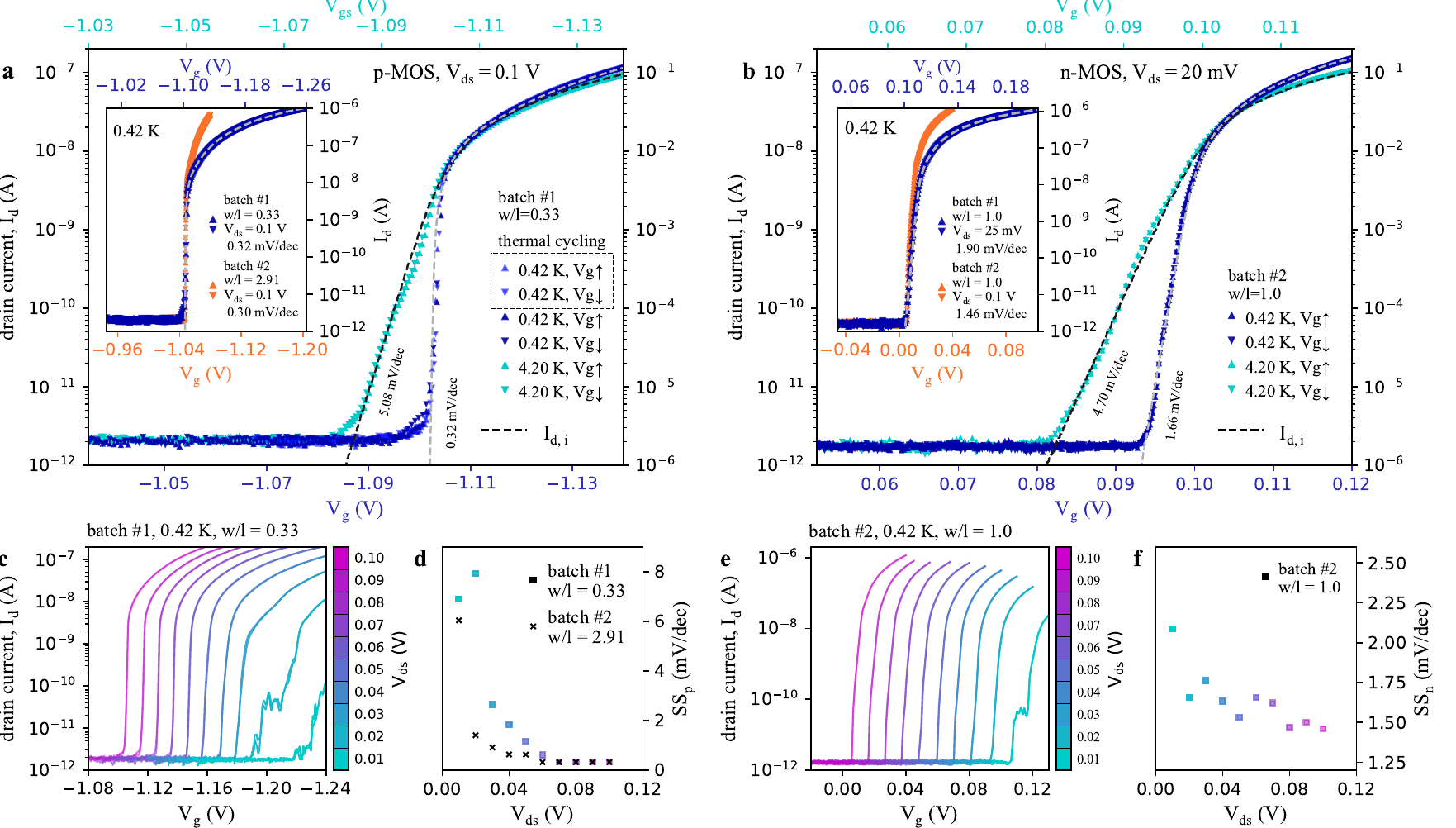}
\caption{Sub-Kelvin FD-SOI MOSFET transport, $I_d(V_g)$ transfer characteristics for a p-MOSFET (\textbf{a}, $V_{ds}=100$ mV), and an n-MOSFET (\textbf{b}, $V_{ds}=20$ mV) at 420~mK and 4.2~K. The arrows in the legend indicate the sweep direction of $V_g$. Inset of a: $I_d(V_g)$ of two p-MOSFETs from different wafer batches with different channel width to gate length ratio w/l. Inset of b: $I_d(V_g)$ of two n-MOSFETs from different wafer batches with the same w/l but different $V_{ds}$ magnitude. \textbf{c,e} $I_d(V_g)$ transfer characteristics are measured at 420 mK at variable $V_\text{ds}$. \textbf{d,f} $SS(V_\text{ds})$ at 420 mK extracted by fitting $I_d(V_g)$ in the subthreshold region.}
\end{figure*}
\noindent MOSFETs were manufactured by SemiQon Technologies on 150 mm SOI wafers. In these, charge carriers are extended from doped contact regions into a 20 nm thick, intrinsic channel, tuneable by front- and back-gate. The front-gate effective gate oxide thickness is 20 nm. The back-gate can be used to tune the threshold voltage so that the channel potential is close to the conduction or valence band edge when the front gate voltage is at zero. In this experiment the back-gate was kept grounded due to pin limitations and the fact that we never found an effect of variable back-gate potential on SS before. 

\noindent Cryogenic characterization was performed with $^3$He evaporation cooling in a cryogen free system (Bluefors LD250HE) with 420 mK base temperature. Source and gate potentials were controlled using isolated ultra-low noise high-stability voltage sources (Quantum Machines, QDAC-II). Source and drain current was measured with a low-noise high-stability transimpedance amplifier (Basel instruments SP983c). The input JFET of the preamplifier has a leakage current of about 2 pA, setting the current floor in all shown $I_d(V_g)$-sets. \\

\noindent To quantify switching metrics of our MOSFETs we sweep $V_g$ and measure $I_d$ at temperature $T$ and source-drain voltage $V_{ds}$. The gate leakage was measured prior to these measurements to be below the noise floor. We extract $SS$ by fitting the log-linear part of $I_d(V_g)$, and use a phenomenological approach to fit $I_d(V_g)$ over the full interval of $V_g$. For this we calculate the carrier density of carrier type $i=n,p$ with
\begin{align}
n_i(\Psi_S) = \int\limits_{-\infty}^{\infty} f(E)N_i^{2D}(E,\Psi_S)\, \mathrm{d} E.
\end{align}
$\Psi_S$ is the surface potential, $f(E)=(\text{e}^{(E-E_F)/k_{B}T}+1)^{-1}$ is the Fermi-Dirac distribution, $N_i^{2D}$ is the two-dimensional density of states which is constant above the band edge $E_i = E_i^0\pm e\Psi_S$. $V_g$ is the sum of the voltage drop over the gate oxide and the semiconductor as \cite{Lundstrom}
\begin{align}
V_{g,i} = \Psi_S + n_i(\Psi_S)e/C_\text{ox}.
\end{align}
$C_\text{ox}=\epsilon/t$ is the geometric gate capacitance where $\epsilon$ is the static permittivity and $t$ is the thickness of the gate dielectric. As in \cite{8660508,Oka2023} we account for subgap states by adding a broadened band edge $N_i(E)=N_i^{2D}\text{e}^{(E\pm E_i)/E_{w,i}}$ for $E<E_i$, with the decay energy $E_{w,i}$ \cite{Oka2023}. We fit measured drain current by using the most trivial approach
\begin{align}
I_{d,i} = wq_iv_i.
\end{align}
with $q_i=\pm en_i$, the width of the channel $w$, and the average charge carrier velocity $v_i$. We assume that (ballistic) mobility is constant in the subthreshold regime, $dn_i/\psi_S = dI_d/dV_g$, fit a normalized subthreshold carrier concentration $n_i$ to $I_d$, and then use $v_i$ as fitting parameter above $V_{th}$. The normalization of $n_i$ creates an ambiguity which makes physical parameter extraction impossible, but allows to estimate relative changes in (above threshold) on-currents with temperature without a density, but rather a transport related parameter.\\

\noindent We benchmark the switching performance of our MOSFETs in a range from 280 K to 420 mK. Results for a typical p-MOSFET and an n-MOSFET are depicted in Fig. 2. In Fig. 2, a and b we plot selected $I_d(V_g-V_\mathrm{th})$ at 280 K, liquid N$_2$ temperature 77 K, and 420 mK to showcase the significant change in turn-on when cooling over this wide temperature range. $V_\mathrm{th}$ is the threshold voltage with $I_d(V_g\leq V_\text{th})\leq 10^{-8}$ A. By fitting the log-linear subthreshold current we find the expected linear decrease in SS with decreasing temperature, $SS(T) = s_i\mathrm{ln}(10) k_{\text{B}} T/e$, with $s_p=1.16$ down to about 100 K, and $s_n=1.15$ down to about 150 K (Fig. 2, d and f). At lower temperatures the measured $SS$ does not follow a linear temperature dependence but plateaus between 10 K and 20 K. Notably, by cooling further we find $SS$ further decreasing linearly for both electrons and holes, reaching 0.32 mV/dec at 420 mK for the p-MOSFET, while forming a second plateau between 1-2 mV/dec for the n-MOSFET below 2 K. At 4.2 K we find 5.1 mV/dec for the p-MOSFET and 4.5 mV/dec for the n-MOSFET (Fig. 2, d and f), well reproducing previous results from different wafer batches, found to be consistent in gate length from 50~nm to >1~\si{\micro\nothing}m \cite{bohuslavskyi2024scalableonchipmultiplexingsilicon}. Regarding the form of $SS_p(T)$ at all $T$, and $SS_n(T)$ above 3 K, our results are in agreement with the findings in \cite{Oka2023}.\\
\noindent Besides an improving $SS$, the on-current for both MOSFETs increases by a decade when cooling from 280 K to 420 mK (Fig. 2, a and b). Fig. 2 c and e display these changes quantified by $v_i$ (see Eq. 3). In Fig. 2, e we compare $v_n$ at 420 mK to the measured electron mobility $\mu_H$ of an n-Hall-bar MOS-test structure with identical $C_\text{ox}$ at 420 mK. We note that the increase in electron mobility above $V_{th}$ qualitatively reproduces the increase in $v_n$ over the range of $V_g$.\\
\noindent At 420 mK we evaluate the switching behaviour of our MOSFETs with respect to hysteresis or transport features originating from single-electron tunneling \cite{9777765} or delocalization of charge carriers \cite{Shashkin-2019}. For this we measure $I_d(V_g,V_\mathrm{ds})$ characteristics of our MOSFETs by sweeping $V_g$ back and forth at variable $V_\text{ds}$ (Fig. 3, c and e). As reference we characterize a device from another wafer batch, with the same w/l (n-MOS), and a 9 times larger w/l (p-MOS). For all devices we find no hysteresis, current jumps, or current oscillations down to $V_\text{ds}=20$~mV, except only the p-MOSFET with w/l = 0.33 showing oscillatory current modulation at $V_{ds}=20$~mV. We find very similar switching behaviour between the devices from different wafer batches, with 1.66~mV/dec at $V_{ds}=20$~mV versus 1.90~mV/dec at $V_{ds}=25$~mV for the n-MOSFETs (Fig. 2, b versus Fig. 3, b), and 0.32 mV/dec versus 0.30 mV/dec for the p-MOSFETs (Fig. 3, a, inset). In the latter, the expected increase in on-current due to the increased w/l can be seen as well. We further probe the sub-1~mV/dec hole turn-on by doing full thermal cycling which does not result in a change of subthreshold swing or threshold voltage (Fig. 3, a). We note that for both electrons and holes we achieve $SS$ below the lowest reported value of 3.5 mV/dec at 420 mK \cite{Oka2023}.\\

\noindent The very sharp turn-on of our MOSFETs enables lowering the supply voltage $V_\mathrm{dd}$ in cryo-CMOS ICs well below that of state-of-the-art FD-SOI, which is currently above 0.4 V \cite{GlobalFoundries2024} and typically above 1 V. When switching from $I_\mathrm{off} = I_\mathrm{d}(V_{g} = 0, V_\mathrm{ds} = V_\mathrm{d})$ to $I_\mathrm{on} = I_\mathrm{d}(V_{g} = V_\mathrm{ds} = V_\mathrm{dd})$ over several decades in current, our MOSFETs can be switched on with $V_\text{dd}$ of the order of tens of millivolts leading to reduction of dynamic power-loss by more than a factor of hundred compared to the state-of-the-art.\\
\noindent In order to discuss the expected power loss of a CMOS IC operating with millikelvin MOSFETs, we depict an inverter that is designed according to the process specifications given by the MOSFET behavior displayed in this work. The inverter is driven by a square wave at frequency $f$ and amplitude $V_\mathrm{dd}$. We set the gate length $Lg=50$ nm and normalize the calculated power to n- and p- specific unit channel widths $W_g$ of 50 nm (n-MOSFET) and 150 nm (p-MOSFET) to match their on-state resistance. The power consumption is then
\begin{align}
P = P_{d} + P_0 = C_\text{t}V_\mathrm{dd}^2f + V_{dd}I_{off} ,
\end{align}
where $C_t = C_i + C_l$ is the total capacitance consisting of the inverter $C_i = 4C_g$ and the fan-out $C_l = 16C_g$, with $C_g=C_{ox}L_gW_g$, defining the dynamic power consumption $P_d$. The static power consumption $P_0$ is dominated by the subthreshold leakage current $I_{off}$ to the ground. At low temperature, we choose $V_{th}$ sufficiently far in the gap and neglect $P_0$ due to the exponential decrease of $I_{off}$ with temperature. In this simple picture, we can estimate realistic limits for $V_{dd}$, when trying to tune the IC supplies towards minimum power. We show the dependence of key power consumption metrics on $V_{dd}$ in Fig. 4.\\
\noindent At lower supply voltages, the sharp turn into the subthreshold region imposes a distinct frequency cutoff. Here, the finite delay in charging and discharging limits the output voltage response $V_\mathrm{out} = V_{ds}(1-e^{-t/R_iC_\mathrm{tot}})$, where $R_i(V_\mathrm{dd})$ is the n- or p-MOSFET resistance. This gives the maximum operation frequency to $f_{max} = 1/\tau_R + 1/\tau_F$, where $\tau_R$ is the rise and $\tau_F$ the fall time. For the values of $f_{max}$ depicted in Fig. 4 We have neglected inductive loads and assumed that the circuit operates between 10 $\%$ (off) and 90 $\%$ (on) $V_\mathrm{dd}$.\\
\noindent A second, separate lower-limit of $V_{dd}$ is defined by the onset of single-electron tunneling effects, as seen in Figs. 3c and 3e, and that is below 30 mV for the MOSFETs used in the experiment. It should be noted that this not an universal value but the threshold depends on the exact nature of confinement potentials.\\
\noindent When increasing $V_{dd}$ further, the diminishing returns on RF performance compared with the increasing power consumption imposes the upper limit of supply voltage, defined by the direct cooling power of the cryostat on the electronic system, which has to be balanced with the total power consumption of the IC. The cooling power of a conventional dilution refrigerator at 50~mK is as high as 100~\si{\micro\nothing}W, while $^3$He evaporation or dilution refrigeration at 500~mK can provide a cooling power of several tens of mW. Comparing this to the power metrics in Fig. 4 clearly shows that our MOSFETs can be incorporated in millikelvin ICs without causing significant thermal overhead. For example, a whole chip with mixed types of ICs consisting of a total of a million MOSFETs having common supply of $V_{dd}=50$ mV and driven with a clock of 10 MHz with all transistors switching on every cycle, would have power loss less than 100 nW. This is many decades less than that is typically reported for cryo-CMOS circuits \cite{Bartee2025}. An analog circuit consisting of 100 transistors would only dissipate 1 pW/MHz, which would still be very small compared to the reduced effective cooling power to such an IC over various types of thermal boundaries.

\begin{figure}
\includegraphics[clip,trim=0.5cm 0.15cm -1cm 1cm,width=0.46\textwidth]{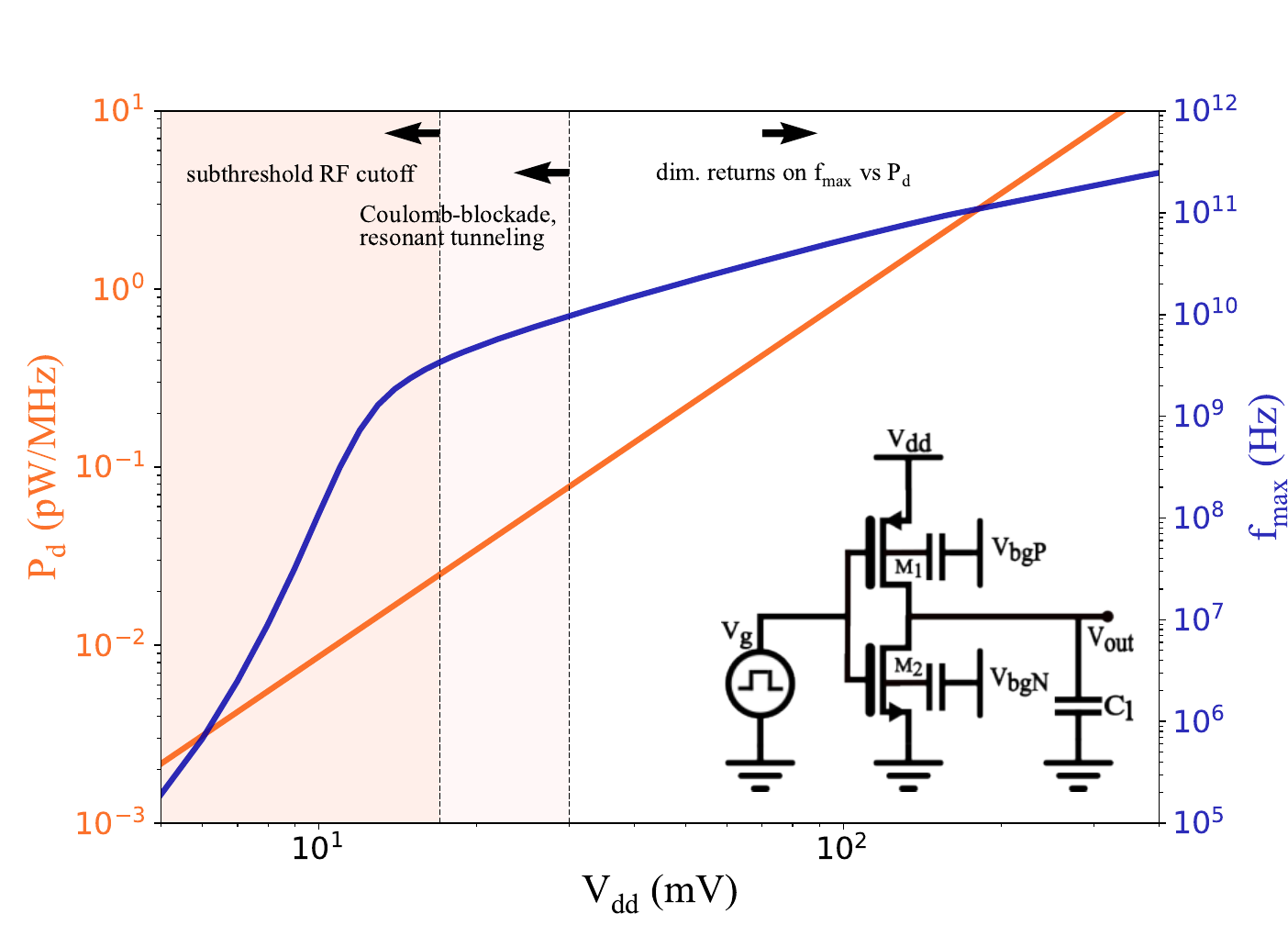}
\caption{Dynamic power consumption per unit frequency in MHz and limit frequency $f_{max}$ of an inverter with fan-out of 4 inverters. Both n-MOSFETs and p-MOSFETs have a gate length of 50 nm and the power displayed on the left axis is the normalized power per different unit widths 50 nm (n-MOSFET) and 150 nm (p-MOSFET) to match their on-state resistance.}
\end{figure}
\noindent In summary, we have characterized cryogenic FD-SOI MOSFETs produced on a CMOS pilot-line and shown that the improvement in switching metrics, due to reduced thermal excitations of charge carriers, can be harnessed down to the millikelvin temperature regime. This advancement can solve the power dissipation bottleneck, limiting very large-scale integration of CMOS circuits at ultra-low temperatures, and enable all-cryogenic control of quantum circuits, which is a prerequisite for fault-tolerant quantum computation.\\

\noindent Acknowledgements: We acknowledge funding by the European Union’s Horizon EIC programme under grant agreement No. 101136793 (SCALLOP) and Chips Joint Undertaking programme under grant agreement No. 101139908 (ARCTIC), European Research Council (ERC) under the European Union’s Horizon 2020 research and innovation program (Grant Agreement No. 852428) and Business Finland project 10200/31/2022 (ToScaleQC). 
 
\bibliography{references}

\end{document}